\documentclass[10pt,aps,prd,notitlepage,preprintnumbers]{revtex4-1}
\pdfoutput=1
\usepackage{graphicx}
\usepackage{slashed}
\usepackage{xcolor}
\usepackage{amsmath}
\usepackage{amssymb}
\usepackage{amsfonts}
\usepackage[pdfborder={0 0 0}]{hyperref}
\hypersetup{
  pdfauthor={Guang-Peng Zhang},
  pdftitle={Probing transverse momentum dependent gluon distribution functions from hadronic quarkonium pair production},
  pdfkeywords={NRQCD, TMD, gluon distribution}
}
\newcommand{\ep}{\epsilon}
\newcommand{\la}{\lambda}
\newcommand{\de}{\delta}
\newcommand{\ga}{\gamma}
\newcommand{\al}{\alpha}
\newcommand{\be}{\beta}
\newcommand{\sig}{\sigma}
\newcommand{\s}[1]{\slashed{#1}}
\newcommand{\no}{\notag\\}
\begin{document}
\title{Probing transverse momentum dependent gluon distribution functions from hadronic quarkonium pair production}
\author{Guang-Peng \surname{Zhang}}
\affiliation{Center for High Energy Physics, Peking University, Beijing 100871, China}

\begin{abstract}
The inclusive hadronic production of $\eta_Q$($\eta_c$ or $\eta_b$) pair is proposed to extract the transverse momentum dependent(TMD) gluon distribution functions.
We use nonrelativistic QCD(NRQCD) for the production of $\eta_Q$. Under nonrelativistic limit TMD factorization for this process is assumed to make
a lowest order calculation.
For unpolarized initial hadrons, unpolarized and linearly polarized gluon distributions can be extracted by studying different angular distributions.
\end{abstract}

\maketitle


\section{Introduction}
Transverse momentum dependent(TMD) parton distribution functions can provide valuable information about the transverse motion of partons in the hadron.
The corresponding factorization theorem, i.e. TMD factorization, has been proven for several processes, e.g. semi-inclusive
deep inelastic scattering\cite{Ji:2004wu}, Drell-Yan processes\cite{Ji:2004xq,Collins:1984kg}. From these processes, one can extract TMD quark distributions(see \cite{Barone:2010zz,D'Alesio:2007jt} for a review). But since
gluon distributions have no contribution in these processes, very little information about them is known. However, the TMD gluon distributions paly an important role at high energy hadron colliders. Especially in \cite{Boer:2011kf,Boer:2013fca}, the authors found the transverse momentum dependence of
Higgs particle production can help to determine the spin and parity of Higgs particle. Several processes have been proposed to extract TMD gluon distributions, such as $A+B\rightarrow\ga+\ga+X$\cite{Qiu:2011ai}, $A+B\rightarrow\eta_c(\eta_b)+X$\cite{Boer:2012bt} and $A+B\rightarrow\ga+J/\psi(\Upsilon)+X$\cite{Dunnen:2014eta}, where the initial hadrons $A,B$ can be proton or antiproton, and $X$ represents all possible final hadrons which are not observed.
However, for the relative complicated production mechanisms, the complete proof of TMD factorization for these processes have not been obtained. In\cite{Ma:2012hh} the factorization for $\eta_c(\eta_b)$ inclusive production is confirmed up to one loop level under nonrelativistic limit. For photon pair production, isolation conditions for the photons may be necessary to exclude the fragmentation production of photons which spoils
TMD factorization. For photon-quarkonium associated production, it is expected that the factorization holds under nonrelativistic limit, for detailed discussion one can consult\cite{Dunnen:2014eta}.

In this work we propose another process to probe TMD gluon distribution functions, i.e. the inclusive $\eta_Q$($\eta_c$ or $\eta_b$) pair production $A+B\rightarrow\eta_Q+\eta_Q+X$ with the two $\eta_Q$'s nearly back-to-back. As done in\cite{Boer:2012bt,Ma:2012hh}, we will use NRQCD to deal with the production of $\eta_Q$ pair and take the nonrelativistic limit. In NRQCD, the heavy quark pair in the quarkonium are nearly
on-shell and have a small relative momentum of order $Mv$, where $v\ll 1$ is the typical velocity of the heavy quark in the rest frame of quarkonium. In general, the quark pair can be in color singlet or octet. It is obvious that the heavy quark pair can only be generated from gluon annihilation at hadron colliders. Thus it is natural to extract gluon distributions from heavy quarkonium production. The main component of $\eta_Q$ is a
colorless heavy quark pair with quantum number $^{2S+1}L_J=^1S_0$, only one matrix element $\langle 0|\mathcal{O}_1^\eta(^1S_0)|0\rangle$ contributes in leading order of $v$. From the argument in \cite{Bodwin:1994jh}, the infrared divergences associated with a soft gluon attached to
the on-shell heavy quarks will cancel out after summing up all diagrams. The exception is Coulomb singularity, which can be absorbed into the NRQCD matrix element. Then all soft divergences can be absorbed by TMD distribution functions or a general soft factor\cite{Ji:2004wu,Ji:2004xq}. Thus TMD factorization is expected for this process.

At leading order the corresponding hard process is $g+g\rightarrow \eta_Q\eta_Q$, another contribution from quark and anti-quark annihilation is negligible at high energy colliders due to the large size of gluon distribution functions. In this work, we consider unpolarized initial hadrons, there are two TMD gluon distributions $f_1^g$ and $h_1^{\perp g}$\cite{Mulders:2000sh}, which represent unpolarized and linearly polarized gluon distributions in the unpolarized hadron, respectively. Both can be extracted from different angular distributions of the quarkonium pair. Note that each quarkonium can have large transverse momentum as long as the total transverse momentum of the quarkonium pair is small. The detection of $\eta_Q$ may be a problem, in this work we are unable to solve this problem. However, as suggested in\cite{Boer:2012bt,Barsuk:2012ic}, $p\bar p$ and $\phi\phi$-channels can be used to detect $\eta_c$. To indicate the possibility of observation of $\eta_c$, we make an estimate about the production cross section based on the tree level calculation.

The organization of
this paper is as follows: in Sec.2 we introduce our notations and formalism; in Sec.3 we calculate the cross section for this process; in Sec.4 we present our numerical results and make a discussion; Sec.5 is our summary.

\section{Kinematics and formalism}
The process we consider is
\begin{align}
A(P_A)+B(P_B)\rightarrow Q\bar{Q}[^1S_0](p_1)+Q\bar{Q}[^1S_0](p_2)+X\rightarrow \eta_Q(p_1)+\eta_Q(p_2)+X.\notag
\end{align}
A,B are two unpolarized hadrons with momenta $P_A$ and $P_B$, respectively. $Q$ is the heavy flavor quark which can be $b$ or $c$ quark in our case. The calculation is performed in center-of-mass(cm) frame, where $P_A $ is along z-axis. For convenience, we choose light-cone coordinates, that is,
$a^\mu=(a^+,\ a^-,\ a_\perp^\mu)$, $a^\pm=(a^0\pm a^3)/\sqrt{2}$ for any four vector $a^\mu$, the transverse direction is relative to the z-axis.
In this work we focus on small $q_\perp$ region, in which the cross section is sensitive to the transverse motion of partons. Of course $Q^2$ should be large
enough to make the perturbative calculation reliable. Hence, we are interested in the following region:
\begin{align}
Q^2\equiv q^2=(p_1+p_2)^2\gg \Lambda_{QCD}^2,\ \ q_\perp^\mu\sim\Lambda_{QCD},
\end{align}
in this region we can make the power expansion in $q_\perp/Q$ and take only leading power or leading twist contribution in this work.
The following invariants will be useful,
\begin{align}
S=(P_A+P_B)^2,\ \ x_a=\frac{q^2}{2P_A\cdot q},\ \ x_b=\frac{q^2}{2P_B\cdot q}.
\end{align}
Furthermore, $x_{a,b}$ can be expressed through the rapidity of the quarkonium pair $y$,
\begin{align}
x_a=\frac{Q}{\sqrt{S}}e^y,\ \ x_b=\frac{Q}{\sqrt{S}}e^{-y},\ \ y\equiv\frac{1}{2}\ln\frac{q^+}{q^-}.
\end{align}
Here we have taken the mass of initial hadrons to be zero under high energy limit.

For the quarkonium, we take nonrelativistic limit, the heavy quark and heavy anti-quark in the quarkonium are on-shell and have the same momentum. For $\eta_Q$ with quantum number $J^{PC}=0^{-+}$, the main component is the colorless
quark pair in the angular momentum state $^1S_0$, the projector for this partial wave is
\begin{align}
\Pi_0(k)=\frac{1}{\sqrt{8M^3}}(-\s{k}+M)\ga_5(\s{k}+M),
\end{align}
where $k^\mu$ is the momentum of quark, $M$ is heavy quark mass. This is the projector in \cite{Petrelli:1997ge}, except for a color factor $1/\sqrt{N_c}$.
The transition rate from $Q\bar Q$-pair to $\eta_Q$ is represented by the NRQCD matrix element
$\langle \mathcal{O}_1^\eta(^1S_0)\rangle\equiv \langle 0|\mathcal{O}_1^\eta(^1S_0)|0\rangle$.

We measure the direction of final quarkonium in the Collins-Soper(CS) frame, which is the rest frame of
quarkonium pair and obtained by two boosts\cite{Collins:1977iv,Arnold:2008kf}. In CS frame, the momentum of one quarkonium $p_1$ is
\begin{align}
p^\mu_1=\tilde{p}_1(\frac{Q}{2\tilde{p}_1},\sin\theta\cos\phi,\sin\theta\sin\phi,\cos\theta),\ \  \tilde{p}_1=\frac{Q}{2}\rho,\ \rho=\sqrt{1-\frac{4M_\eta^2}{Q^2}}.
\end{align}

\section{Cross section}
\begin{figure}
\begin{center}
\includegraphics[width=.3\textwidth]{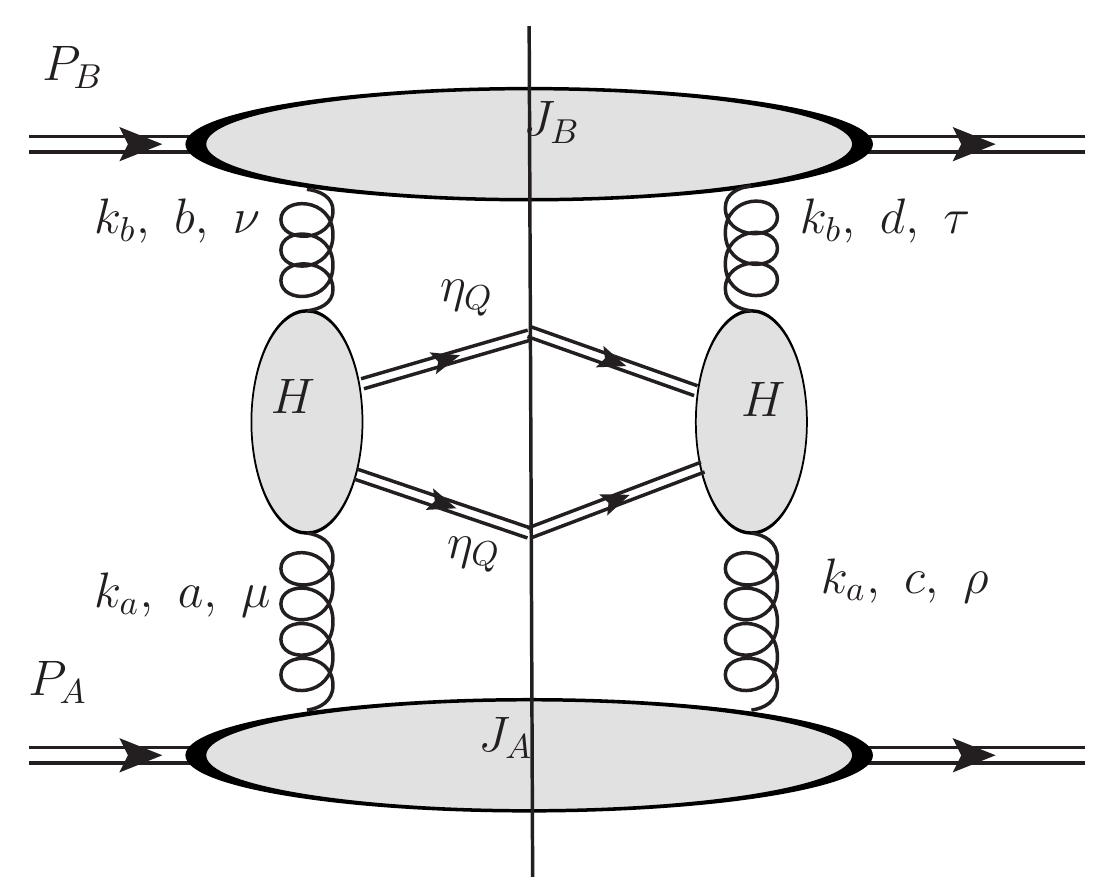}
\end{center}
\caption{The general diagram for pair $\eta_c$ production. The central bubble represents the hard part, in which all propagators are far off-shell.
$abcd$ and $\mu\nu\rho\tau$ are color and Lorentz indices for the partons(gluons).}\label{pinch-surf}
\end{figure}
Assuming that TMD factorization holds, the cross section can be factorized into the convolution of a hard scattering part and TMD gluon distributions,
as shown in Fig.\ref{pinch-surf}. At leading order of $\al_s$, the subprocess is $g(k_a)g(k_b)\rightarrow \eta_Q(p_1)\eta_Q(p_2)$. At leading twist, the gluons going into the hard scattering are collinear to hadrons $A,\ B$, that is,
\begin{align}
k_a^\mu=(k_a^+,\ k_a^-,\ k_{a\perp}^\mu)\sim Q(1,\la^2,\la),\ \ k_b^\mu=(k_b^+,\ k_b^-,\ k_{b\perp}^\mu)\sim Q(\la^2,1,\la),\ \la\simeq \Lambda_{QCD}/Q.
\end{align}
At the leading order $k_a^-$ and $k_b^+$ can be set to zero in the hard part. For $k_{a,b\perp}^\mu$, since they are of the same order as $q_\perp$, we should
retain $\de^2(q_\perp-k_{a\perp}-k_{b\perp})$, but for the hard scattering amplitude, both $k_{a,b\perp}^\mu$ can be set to zero. Then the
cross section is
\begin{align}
d\sig=&\frac{(2\pi)^4}{S^2}\frac{\de^{ac}\de^{bd}}{N_c^2(N_c^2-1)^2}\frac{1}{2}d\Phi_2\int_\perp \Phi_A^{\mu\rho}(x_a,k_{a\perp})
\Phi_B^{\nu\tau}(x_b,k_{b\perp})\mathcal{M}^{ab}_{\mu\nu}(\mathcal{M}^{cd}_{\rho\tau})^*\langle \mathcal{O}_1^\eta(^1S_0)\rangle^2,\no
\int_\perp=&\int d^2k_{a\perp}d^2k_{b\perp}\de^2(k_{a\perp}+k_{b\perp}-q_\perp),
\label{cross-section}
\end{align}
where $d\Phi_2=d^3 p_1 d^3 p_2/(2\pi)^6(4E_1E_2)$ is the phase space integration measure. $\mathcal{M}^{ab}_{\mu\nu}$ is the amplitude for the hard scattering
$gg\rightarrow Q\bar Q+Q\bar Q$ projected to $^1S_0$ partial wave, without the polarization vectors for external gluons.
This cross section formula can be obtained from the standard procedure, e.g., \cite{Collins:2008sg}.
For unpolarized hadrons, at leading twist we can define two TMD gluon distributions $f_1^g,\ h_1^{\perp g}$, which represent the unpolarized and linearly polarized gluon distributions in the hadron\cite{Mulders:2000sh},
\begin{align}
\Phi_A^{\mu\rho}(x_a,k_\perp)=&\int\frac{d\xi\cdot P_A d^2\xi_\perp}{(x_a P_A^+)^2(2\pi)^3}e^{i(x_a P_A\cdot\xi+k_\perp\cdot\xi_\perp)}
\langle P_A|G^{+\rho}_a(0)G^{+\mu}_a(\xi^-,\xi_\perp)|P_A\rangle|_{\xi^+=0}\no
=&-\frac{1}{2x_a}(g_\perp^{\mu\rho}f_1^g(x_a,k_{\perp}^2)-\frac{1}{2m_A^2}k_{\perp}^{\{\mu}k_\perp^{\rho\}}h_1^{\perp g}(x_a,k_{\perp}^2)).
\label{TMD-def}
\end{align}
where $m_A$ is the hadron mass, $\mu\rho$ are transverse and $a_\perp^{\{\mu}b_\perp^{\nu\}}\equiv a_\perp^{\mu}b_\perp^\nu+a_\perp^\nu b_\perp^\mu-g_\perp^{\mu\nu}a_\perp\cdot b_\perp$.
In a similar way one can define $\Phi_B^{\nu\tau}$. For simplicity, the Wilson lines are suppressed.

Now we turn to calculate the amplitude $\mathcal{M}^{ab}_{\mu\nu}$ in the cross section eq.(\ref{cross-section}). As mentioned above, all external momenta are
on shell, and $p_{2\perp}=-p_{1\perp}$ since we demand $q_\perp\rightarrow 0$, there is only one independent transverse momentum in the amplitude.
Considering P-parity invariance and colorless condition, the scattering amplitude can be decomposed as
\begin{align}
\label{scalarM}
i\mathcal{M}_{ab}^{\mu\nu}=&\de^{ab}\left(\frac{p_{1\perp}^\mu p_{1\perp}^\nu}{p_{1\perp}\cdot p_{1\perp}}M_1+g_\perp^{\mu\nu}M_2 \right),
\end{align}
and $M_{1,2}=M_{1,2}(s,t,u)$ is symmetric in $t,u$, where $s=(k_a+k_b)^2$, $t=(k_a-p_1)^2$, $u=(k_a-p_2)^2$.

\begin{figure}
\begin{flushleft}
\begin{minipage}{0.18\textwidth}
\includegraphics[scale=0.3]{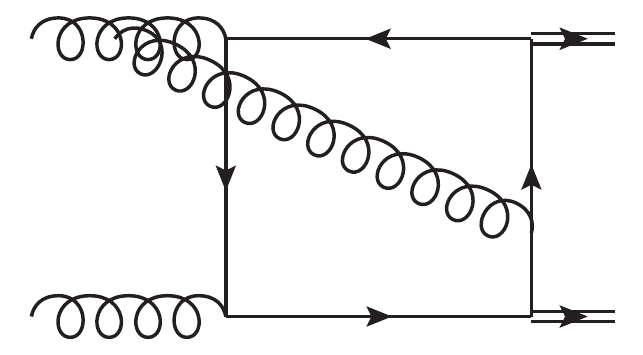}\\
(a)
\end{minipage}
\begin{minipage}{0.18\textwidth}
\includegraphics[scale=0.3]{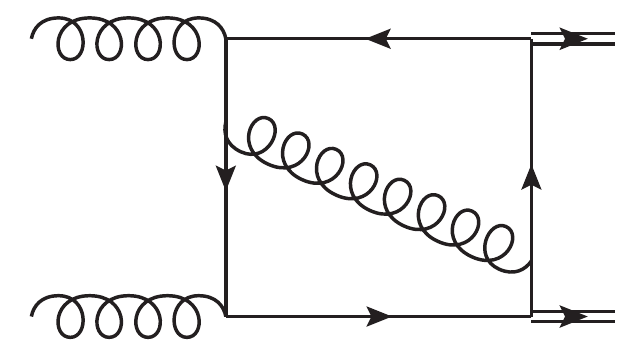}\\
(b)
\end{minipage}
\begin{minipage}{0.18\textwidth}
\includegraphics[scale=0.3]{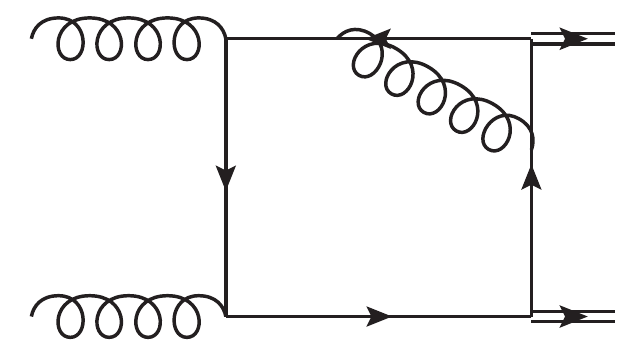}\\
(c)
\end{minipage}
\begin{minipage}{0.18\textwidth}
\includegraphics[scale=0.3]{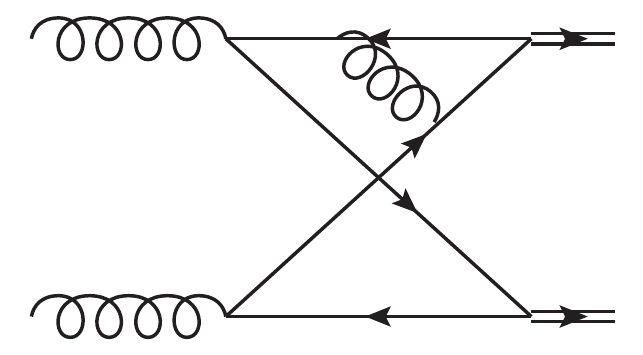}\\
(d)
\end{minipage}
\begin{minipage}{0.18\textwidth}
\includegraphics[scale=0.3]{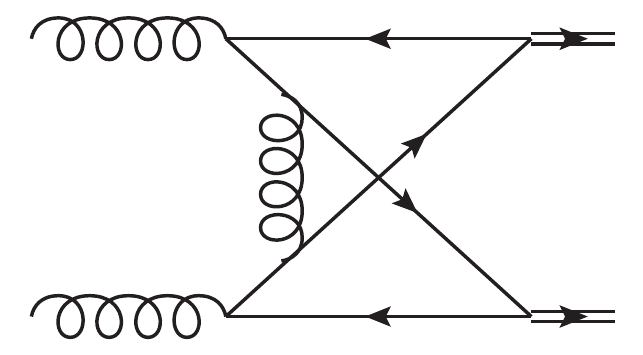}\\
(e)
\end{minipage}
\begin{minipage}{0.18\textwidth}
\includegraphics[scale=0.3]{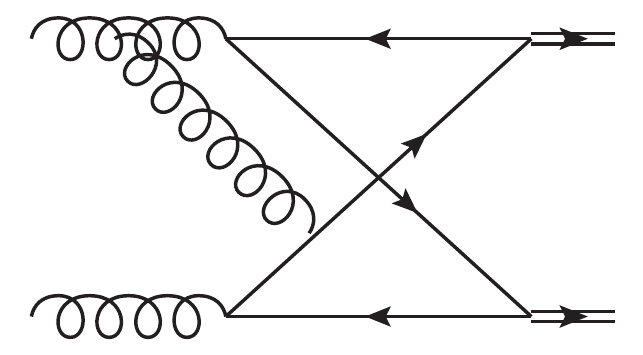}\\
(f)
\end{minipage}
\begin{minipage}{0.18\textwidth}
\includegraphics[scale=0.3]{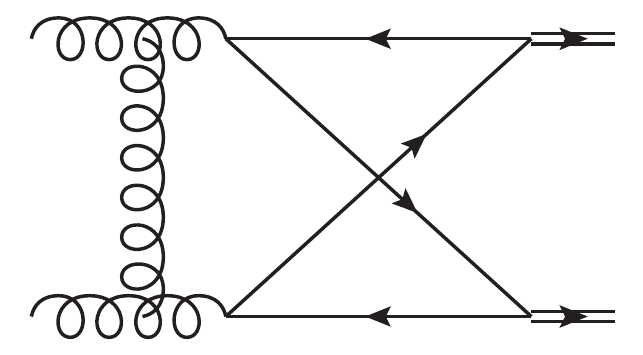}\\
(g)
\end{minipage}
\begin{minipage}{0.18\textwidth}
\includegraphics[scale=0.3]{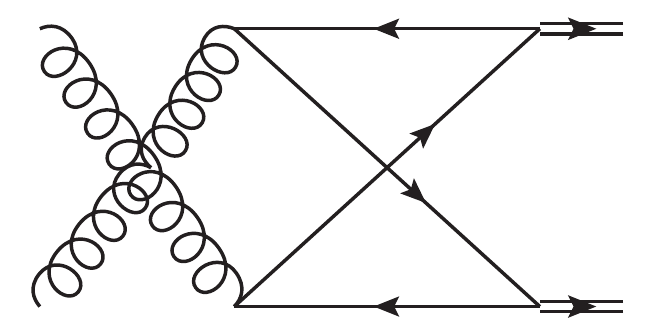}\\
(h)
\end{minipage}
\begin{minipage}{0.18\textwidth}
\includegraphics[scale=0.3]{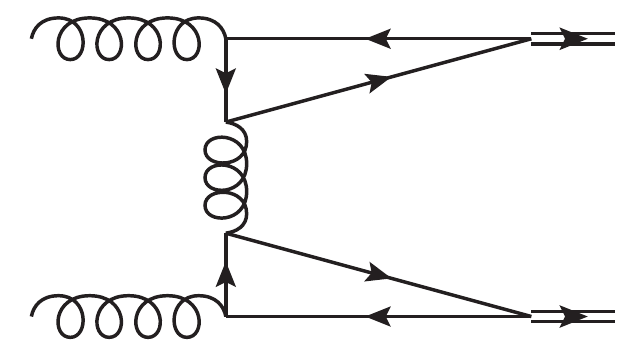}\\
(i)
\end{minipage}
\end{flushleft}
\caption{Some diagrams for the sub-process $g(k_a)g(k_b)\rightarrow \eta_Q(p_1)\eta_{Q}(p_2)$, the double line in the diagrams means heavy quarkonium $\eta_Q$. Other diagrams can be obtained by reversing the directions of fermion loops or exchanging the two initial gluons, some care should be taken to avoid double counting.}\label{sub-diagram}
\end{figure}

All tree level Feynman diagrams are shown in Fig.\ref{sub-diagram}, there are two classes, one contains just a single fermion loop(which is not a true loop, one can understand it as a Dirac trace), the
other contains two fermion loops. To evaluate them, notice that if the difference of two diagrams is just the direction of fermion loop, then the two diagrams
have the same contribution. This is ensured by the charge conjugation symmetry of QCD and the demand of color singlet for $Q\bar Q$ pair. Expand $M_{1,2}$ to the leading order of $\al_s$,
\begin{align}
M_i=\frac{g_s^4}{8M^3}[\tilde{M}_i+\mathcal{O}(\al_s)],\ i=1,2
\end{align}
The result can be written as
\begin{align}
\ \tilde{M}_1=&\frac{4 \left(c^2-1\right) \rho ^2 \left(\rho ^2-1\right)}{3 \left(c^2 \rho ^2-1\right)^2
   \left(\left(2-4 c^2\right) \rho ^2+\rho ^4+1\right)}
   \left(\left(431 c^2-205\right) \rho ^2-\left(9 c^4+c^2\right) \rho^8\right.\no
   &\left.+\left(-437 c^4+205 c^2-98\right) \rho ^4+\left(36 c^6-14 c^4+105
   c^2-1\right) \rho ^6-108\right),\no
\ \tilde{M}_2=&\frac{2 \left(\rho ^2-1\right)}{3
   \left(c^2 \rho ^2-1\right)^2 \left(\left(2-4 c^2\right) \rho ^2+\rho^4+1\right)}
    \left(\left(1858 c^2-623\right) \rho^2+c^4 \left(36 c^2-17\right) \rho ^{10}\right.\no
   &+\left(-2657 c^4+1082
   c^2-295\right) \rho ^4+\left(1400 c^6-695 c^4+586 c^2+15\right) \rho^6\no
   &\left.+\left(-144 c^8+140 c^6-375 c^4+2 c^2\right) \rho ^8-313\right),
\end{align}
where $c=\cos{\theta_{cs}}$, $\rho=\sqrt{1-4M_\eta^2/Q^2}$ and we have taken $N_c=3$.

The following formulas can help us to get the angular distribution. First let us define
\begin{align}
\langle k_{i_1}^{\mu_1}k_{i_2}^{\mu_2}\cdots k_{i_n}^{\mu_n} \rangle\equiv\int_\perp (k_{i_1}^{\mu_1}k_{i_2}^{\mu_2}\cdots k_{i_n}^{\mu_n})f(k_{a\perp}^2, k_{b\perp}^2),
\end{align}
where $k_{i_n}=k_{a\perp}$ or $k_{b\perp}$, $f(k_{a\perp}^2,\ k_{b\perp}^2)$ can be any scalar function.
Then we have
\begin{align}
\langle k_{i\perp}^{\{\mu}k_{i\perp}^{\nu\}}\rangle=&x^{\{\mu}x^{\nu\}}\langle w_i\rangle,\ i=a,b,\no
\langle k_{a\perp}^{\{\mu}k_{a\perp}^{\rho\}} k_{b\perp}^{\{\nu}k_{b\perp}^{\tau\}}\rangle=&
\langle C_1\rangle(x^\mu y^\rho+x^\rho y^\mu)(x^\nu y^\tau+x^\tau y^\nu)+\langle C_3\rangle(x^\mu x^\rho-y^\mu y^\rho)(x^\nu x^\tau-y^\nu y^\tau),\label{Cij}
\end{align}
where
\begin{align}
&w_a=(2(k_{a\perp}\cdot x)^2+k_{a\perp}^2),\ w_b=(2(k_{b\perp}\cdot x)^2+k_{b\perp}^2),\no
&C_1=4x\cdot k_{a\perp}x\cdot k_{b\perp}y\cdot k_{a\perp}y\cdot k_{b\perp},\
C_3=(2x\cdot k_{a\perp}x\cdot k_{a\perp}+k_{a\perp}^2)(2x\cdot k_{b\perp}x\cdot k_{b\perp}+k_{b\perp}^2),
\end{align}
and $x^\mu\equiv{q_\perp^\mu}/{\sqrt{-q_\perp^2}},\ \ y^\mu\equiv \ep^{\mu\nu+-}x_\nu$.

After using eq.(\ref{scalarM},\ref{Cij}) and the definition of TMD gluon distributions in eq.(\ref{TMD-def}), the cross section eq.(\ref{cross-section}) is
\begin{align}
\frac{d\sig}{dydQd^2q_\perp d\Omega}
=&\frac{\pi^2\al_s^4\langle \mathcal{O}_1^\eta(^1S_0)\rangle^2\rho}{N_c^2(N_c^2-1)M_\eta^6 S Q}\int_\perp\left( B_1[f^g_{1A}f^g_{1B}]+B_2\cos{2\phi}[\frac{w_b}{2m_B^2} f^g_{1A}h_{1B}^{\perp g}+\frac{w_a}{2m_A^2}h_{1A}^{\perp g} f^g_{1B}]\right.\no
&\left.+B_3[\frac{C_1+C_3}{4m_A^2 m_B^2}h_{1A}^{\perp g} h_{1B}^{\perp g}]+B_4\cos{4\phi}[\frac{C_1-C_3}{4m_A^2 m_B^2}h_{1A}^{\perp g} h_{1B}^{\perp g}]\right),
\label{main-result}
\end{align}
where $\Omega$ is the solid angle defined in CS-frame for one $\eta_Q$ and $y$ is the rapidity of the $\eta_Q$ pair.
The hard coefficients $B_i$ can be expressed through $\tilde{M}_{1,2}$ as follows. Since at this order $\tilde{M}_{1,2}$ are real, we can take $\tilde{M}^*_{1,2}=\tilde{M}_{1,2}$.
\begin{align}
B_1=&|\tilde{M}_1|^2+(\tilde{M}_1 \tilde{M}_2^*+\tilde{M}_1^* \tilde{M}_2)+2|\tilde{M}_2|^2,\ B_2=|\tilde{M}_1|^2+(\tilde{M}_1 \tilde{M}_2^*+\tilde{M}_1^* \tilde{M}_2),\no
B_3=&\frac{1}{2}|\tilde{M}_1|^2+(\tilde{M}_1 \tilde{M}_2^*+\tilde{M}_1^* \tilde{M}_2)+2|\tilde{M}_2|^2,\ B_4=-\frac{1}{2}|\tilde{M}_1|^2.
\end{align}

\section{Numerical result}
In this section we will use the factorized cross section eq.(\ref{main-result}) to make an estimate. We consider A,B to be protons and $\eta_Q$ to be $\eta_c$. Since the property of TMD gluon distributions are
unknown, we use Gauss model\cite{Schweitzer:2010tt} to parameterize them, as done in\cite{Boer:2012bt,Qiu:2011ai},
\begin{align}
f_1(x,k_\perp)=\frac{1}{\pi\be}e^{-\frac{{\vec k}_\perp^2}{\be}}f_1(x),
\end{align}
where $\be$ is the average transverse momentum of the parton(gluon) in proton. In the model, the $q_\perp$-integrated cross sections are not sensitive to the value of $\be$, here we set $\be=0.5 GeV^2$. $f_1(x)$ is the usual integrated gluon Parton Distribution Function(PDF), we take it as MRSTMCal gluon PDF at $Q^2=100GeV^2$, $f_1(x)=0.75 x^{-1.56}(1-x)^{0.25}$.
The positivity gives a constrain to linearly polarized gluon TMD distribution, i.e.,
$\frac{{\vec k}_\perp^2}{2m^2}|h_1^\perp(x,k_\perp)|\leq f_1(x,k_\perp)$\cite{Mulders:2000sh}.
In order to obtain the maximum of cross section, the positivity bound saturation is
assumed,
\begin{align}
|h_1^\perp(x,k_\perp)|\simeq \frac{2m^2}{\be}f_1(x,k_\perp).
\end{align}
The NRQCD matrix element $\langle \mathcal{O}^\eta_1(^1S_0)\rangle$ can be extracted from the decay width of $\eta_c\rightarrow 2\ga$, at leading order of $v$,
\begin{align}
\Gamma_{\ga\ga}=&(\frac{2}{3})^4\frac{4\pi N_c\al_{em}^2}{M^2}\langle \mathcal{O}^\eta_1(^1S_0)\rangle,
\end{align}
this is the same as \cite{Bodwin:1994jh}, but with the normalization of $\mathcal{O}^\eta_1(^1S_0)$ in \cite{Petrelli:1997ge}, there is a difference of $N_c$ factor between these two formalisms. Then $\langle \eta_c|\mathcal{O}_1(^1S_0)|\eta_c\rangle\simeq 0.02 GeV^3$ which can be taken as the value of $\langle \mathcal{O}^\eta_1(^1S_0)\rangle$, the caused error is of order $v^4$\cite{Bodwin:1994jh}.

Using above parametrization, the weighted differential cross sections are calculated,
\begin{align}
&\langle w(\phi)\rangle\equiv\int dQ d\Omega d^2 q_\perp w(\phi) \frac{d\sig}{dydQd^2q_\perp d\Omega},\label{weighted-cross}
\end{align}
where $w(\phi)$ can be $1,\ \cos{2\phi},\ \cos{4\phi}$.
It is interesting to note that the $\phi$ independent $h_1^{\perp g}h_1^{\perp g}$-term in eq.(\ref{main-result}) has no contribution,
because the coefficient $C_1+C_3$ vanishes after integrating over $q_\perp$.
The results for the three weighted cross sections are summarized in Table.\ref{Num-table}, where $\sqrt{S}=7 TeV$, $y=0$ and $p_{1\perp}$ is constrained to be
larger than 1GeV. If $\sqrt{S}=14 TeV$, the cross sections will increase by $2\sim 3$ times.
\begin{table}
\caption{The weighted differential cross sections obtained from Gaussian model at $\sqrt{S}=7TeV$ and $y=0$, as defined in eq.(\ref{weighted-cross}). In the calculation, we choose $\al_s=0.15$, $M_\eta=3.0GeV$, and ignore all scale dependence.}\label{Num-table}
\tabcolsep0.24in
\begin{tabular}{|c|cccc|}
\hline
\ & $Q(GeV)\in(6.0,10.0)$ & $(10.0,\ 15.0)$& $(15.0,\ 20.0)$& $(20.0,\ 40.0)$\\
\hline
$\langle 1\rangle(pb)$& $2.3\times 10^4$ &$1.7\times 10^3$ & $1.8\times 10^2$& $1.3\times 10^2$\\
$|\langle \cos2\phi\rangle|(pb)$& $2.4\times 10^3$ &$4.6\times 10^2$ & $0.72\times 10^2$& $0.63\times 10^2$\\
$\langle \cos4\phi\rangle(pb)$& $0.20\times 10^2$ &$9.1$ & $2.5$& $3.3$\\
\hline
\end{tabular}
\end{table}

\begin{figure}[h]
\begin{center}
\includegraphics[width=0.5\textwidth]{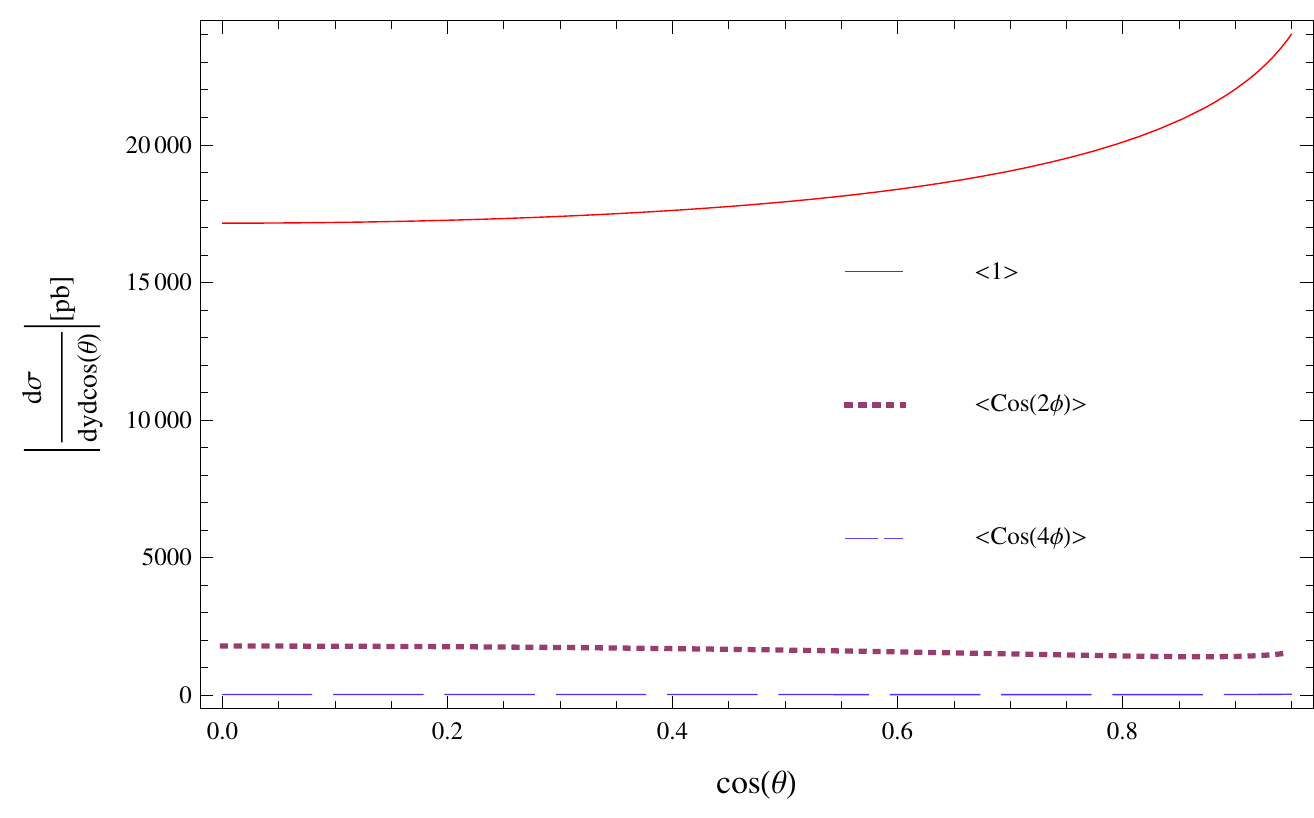}
\end{center}
\caption{The angular dependence for different weighted cross sections, with $y=0$ and $Q$ integrated over $6\sim 40$GeV. We only show the results in the region $\cos{\theta}>0$, since these cross sections are symmetric in $\cos\theta$.}\label{angular-dep}
\end{figure}

Notice that $x_{a,b}=Q/\sqrt{S}$, near threshold $Q=6.0 GeV$ one has $x_{a,b}\sim 10^{-3}$. Due to the large gluon density at small $x$ region, the cross
sections will become relative large near threshold. With the increasing of $Q$, the cross sections will decrease rapidly. It is not expected to get a large
number of events at the region far away from threshold.
Another issue we concern is the $\theta$-dependence, as shown in Fig.\ref{angular-dep}. The curves in the figure represent the differential cross sections $d\sig/{dyd\cos{\theta}}$ weighted with $1,\ \cos{2\phi},\ \cos{4\phi}$, in which $Q$ is integrated over $6\sim 40$GeV and $y=0$. From the result, only for unweighted cross section, i.e. the contribution of unpolarized gluon distribution, has a little enhancement at forward direction. Thus one will not lost most events even excluding the events at forward direction.

But unfortunately, $\eta_c$ is very hard to detect in experiment, even outside the forward direction. In \cite{Barsuk:2012ic} the decay channels $\eta_c$ to $p\bar{p},\ \phi\phi$ are suggested. The detailed discussion about the detection is obviously beyond the scope of our paper. Suppose we can identify $\eta_c$ through $p\bar{p}$ channel, the corresponding branching ratio $(1.51\pm 0.16)\times 10^{-3}$\cite{Beringer:1900zz} means
\begin{align}
&\langle 1,\cos{2\phi}\rangle\times Br^2(\eta_c\rightarrow p\bar{p})\simeq 1\sim 50 fb\notag
\end{align}
at central rapidity region $y=0$, the corresponding events may be observed with the increasing of integrated luminosity at LHC. For $\langle \cos{4\phi}\rangle$, the corresponding value is negligible.

Before ending this section, it is necessary to
mention that we demand the $\eta_c$'s are produced from the hard scattering, rather than the decay of other hadrons. We expect the two cases can be distinguished by proper cut conditions in experiment. Another problem is the relativistic correction which is $v^2$-suppressed relative to the leading power contribution. Notice that for charmonium $v^2\simeq 0.3$\cite{Bodwin:2006dn}, this is actually not very small. Up to order $v^2$, there is only one NRQCD operator $\mathcal{P}_1(^1S_0)$ contributing to the correction\cite{Bodwin:1994jh}, it is interesting to investigate whether the factorization theorem holds in this case. We will study these problems in further work.

\section{Summary}
In this work we propose to use the hadronic production of $\eta_Q$ pair to extract TMD gluon distributions $f_1^g$ and $h_1^{\perp g}$. We work in the framework of NRQCD
and TMD factorization formalism. Under nonrelativistic limit, we expect TMD factorization to hold for this process
since color-octet contribution, which may spoil TMD factorization, is power suppressed by $v^4$. For unpolarized initial hadrons, the resulted cross section has three definite angular distributions, which are proportional to $1,\ \cos{2\phi}$ and $\cos{4\phi}$, respectively. Assuming Gauss model and positivity bound saturation, we make
an estimate for the three angular distributions at LHC with $\sqrt{S}=7TeV$ for $\eta_c$ pair production. The unweighted and $\cos{2\phi}$ weighted cross sections are at $nb$ level
near threshold, and have no obvious enhancement at forward direction, this makes the extraction of TMD gluon distributions $f_1^g$ and $h_1^{\perp g}$ from the two angular distributions possible.

\begin{acknowledgements}
The author would like to thank Y.~Jia for helpful discussion and especially thank J.P.~Ma for a critical reading of the manuscript.
\end{acknowledgements}

\bibliography{reference}

\end{document}